# Programmable photonic signal processor chip for radiofrequency applications

Leimeng Zhuang[1*], Chris G. H. Roeloffzen[2], Marcel Hoekman[3], Klaus-J. Boller[4] and Arthur J. Lowery[1,5]

[1]*Electro-Photonics Laboratory, Electrical and Computer Systems Engineering, Monash University, Clayton, VIC3800, Australia*
[2]*SATRAX BV, PO Box 456, Enschede, 7500 AL, The Netherlands*
[3]*LioniX BV, PO Box 456, Enschede, 7500 AL, The Netherlands*
[4]*Laser Physics and Nonlinear Optics group, University of Twente, PO Box 217, Enschede, 7500 AL, The Netherlands*
[5]*Centre for Ultrahigh-bandwidth Devices for Optical Systems (CUDOS), Australia*
*leimeng.zhuang@monash.edu

**Abstract**
Integrated microwave photonics, an emerging technology combining RF engineering and integrated photonics, has great potential to be adopted for wideband, flexible analog processing applications. However, realizing an application-specific photonic integrated circuit is expensive and time-consuming. Here, we introduce a disruptive approach to tackle this bottleneck, which is analogous to electronic field-programmable gate array (FPGA). We use a grid of tunable Mach-Zehnder couplers interconnected in a two-dimensional mesh network topology, each working as a photonic processing unit. Such a device is able to be programmed into many different circuit architectures and thereby provide a diversity of functions. This paper provides the first-ever demonstration of this concept and show that a programmable chip with a free spectral range of 14 GHz enables RF filters featuring continuous, over-two-octave frequency coverage, i.e. 1.6–6 GHz, and variable passband shaping ranging from a 55-dB-extinction notch filter to a 1.6-GHz-bandwidth flat-top filter.

**Introduction**
Modern radio frequency (RF) systems, such as radio communications, radars, sensor networks, and THz-imaging demand ever-increasing bandwidth and frequency agility[1-3]. At the same time, they require devices that are small, lightweight and low-power, exhibiting large tunability and strong immunity to electromagnetic interference. Integrated microwave photonics[4-7], an emerging technology combining RF engineering and integrated photonics, has the potential to satisfy these needs. Harnessing the large bandwidth and tunability uniquely offered by photonic devices, it enables wideband, flexible front-end analog solutions to precede the digital signal processors that are currently limited to several gigahertz analog bandwidth[8].

In integrated microwave photonics, many key RF functions have been demonstrated using on-chip photonic signal processors, including spectral filters[9-11], phase shifters[12], integrators[13], differentiators[14], pulse shapers[15,16], frequency discriminators[17,18], tunable delay lines[19,20], and beamformers[21,22]. Next to the general advantages of photonic integration in terms of device size, robustness, power efficiency and low-cost potential[23-26], some salient works showed remarkable features such as terahertz processing bandwidths[27], sub-volt control[28], filter extinction ratios greater than 60 dB[29] and multi-octave continuous frequency shifting[19]. Although various and promising, all such demonstrations to date are based on custom designs of an application-specific photonic integrated circuit, the realization of which is expensive and time-consuming. This is a major bottleneck in the development of integrated microwave photonics. An effective solution is to come up with the photonic analog of electronic FPGAs[30], which will save the considerable cost and delay for the chip fabrication.

Here, we propose a novel kind of photonic signal processor chip as is depicted in Fig. 1. The processor features full flexibility in circuit architecture and full control of all circuit parameters in terms of both amplitude and phase. This unique combination is enabled by means of a grid of tunable Mach-Zehnder (MZ) couplers interconnected in a two-dimensional mesh network topology (Fig. 1a), with the MZ couplers being the inter-cell pathways. Each MZ coupler works as a photonic processing unit with freely programmable path-selecting, splitting/combining, and phase shifting capabilities. This makes it possible to define amplitude- and phase-controlled optical routing paths in a two-dimensional plane and thereby create photonic circuits at will, such as the examples shown in Figs. 1b to d. We anticipate this concept to be the starting point of transferring the inestimable enabling power of electrical FPGAs to photonic integrated circuits.

To give an experimental demonstration of this concept, we present here for the first time such a programmable photonic chip. For a proof of principle, the layout comprises a maximally simplified but yet fully versatile 2 × 1 mesh network with two cells. We show that the simple dual-cell circuit with a free spectral range (FSR) of 14 GHz and full parameter-programmability enables RF filters featuring continuous, over-two-octave frequency coverage, i.e. 1.6–6 GHz, and variable passband shaping ranging from a 55 dB-extinction notch filter to a 1.6 GHz-bandwidth flat-top filter. Aiming for a future technology enabler, the results presented here pave the way for the realization of powerful photonic signal processing engines.

**Device principle**
Figure 1a depicts a general waveguide mesh network comprising $M \times N$ square mesh cells, with the MZ couplers being the inter-cell pathways. Each MZ coupler has $2 \times 2$ connection ports, so it can be simultaneously connected to up to four other MZ couplers that constitute the mesh network. We implement the MZ couplers with phase tuning element $\phi_U$ on the upper arm and $\phi_L$ on the lower arm as shown in Fig. 1a. The transfer matrix parameters $c_{ij} = \text{Out}_i / \text{In}_j$ for such a coupler are:

$$c_{11} = -c_{22} = -j\exp(j\phi_A) \cdot \sin(\phi_D) \cdot \exp(j2\pi f \Delta\tau) \quad \text{(bar port)} \quad (1)$$

$$c_{12} = c_{21} = -j\exp(j\phi_A) \cdot \cos(\phi_D) \cdot \exp(j2\pi f \Delta\tau) \quad \text{(cross port)} \quad (2)$$

where $\phi_A = (\phi_U + \phi_L)/2$ and $\phi_D = (\phi_U - \phi_L)/2$ respectively govern the phase and coupling coefficient of its output ports, and $\exp(j2\pi f \Delta\tau)$ represents the frequency-dependent phase shift caused by the propagation delay $\Delta\tau$ of the coupler. By controlling $\phi_D$, an MZ coupler is able to function as an arbitrary-ratio coupler ( $0 < \sin(\phi_D)$, $\cos(\phi_D) < 1$), or function simply as a length of 2-port waveguide with the coupler either in bar-status ($\sin(\phi_D) = 1$ and $\cos(\phi_D) = 0$) or cross-status ($\sin(\phi_D) = 0$ and $\cos(\phi_D) = 1$) as illustrated in Fig. 1a. In the latter function, the cross-bar status of the MZ couplers determines the routing direction of the light from one such waveguide to the next in the mesh network, and the total length of a routing path can be defined by allowing the light to travel through a corresponding number of such waveguides. Figure 1b illustrates the basic circuit building blocks and their implementations in such a mesh network by programming the MZ couplers accordingly, including a $2 \times 2$ coupler, a length of 2-port waveguide, and a circular waveguide loop. Based on this programming mechanism, one can synthesize various circuit architectures in the mesh network. Figure 1c illustrates the implementations of the two general types of waveguide filters that are commonly used to perform signal processing, namely finite impulse response filters based on tapped-delay-lines and infinite impulse response filters based on ring resonators[35]. As far as the mesh network dimension allows, one can reach circuit architectures with arbitrarily extendable functionality, an example of which is depicted in Fig. 1d. It is important to mention that next to the freedom in circuit architectures, $\phi_D$ in the couplers and $\phi_A$ in the constituent waveguides provide the defined circuits with full control capabilities of the amplitude and phase of the light, which facilitates the complete function-programmability of the device.

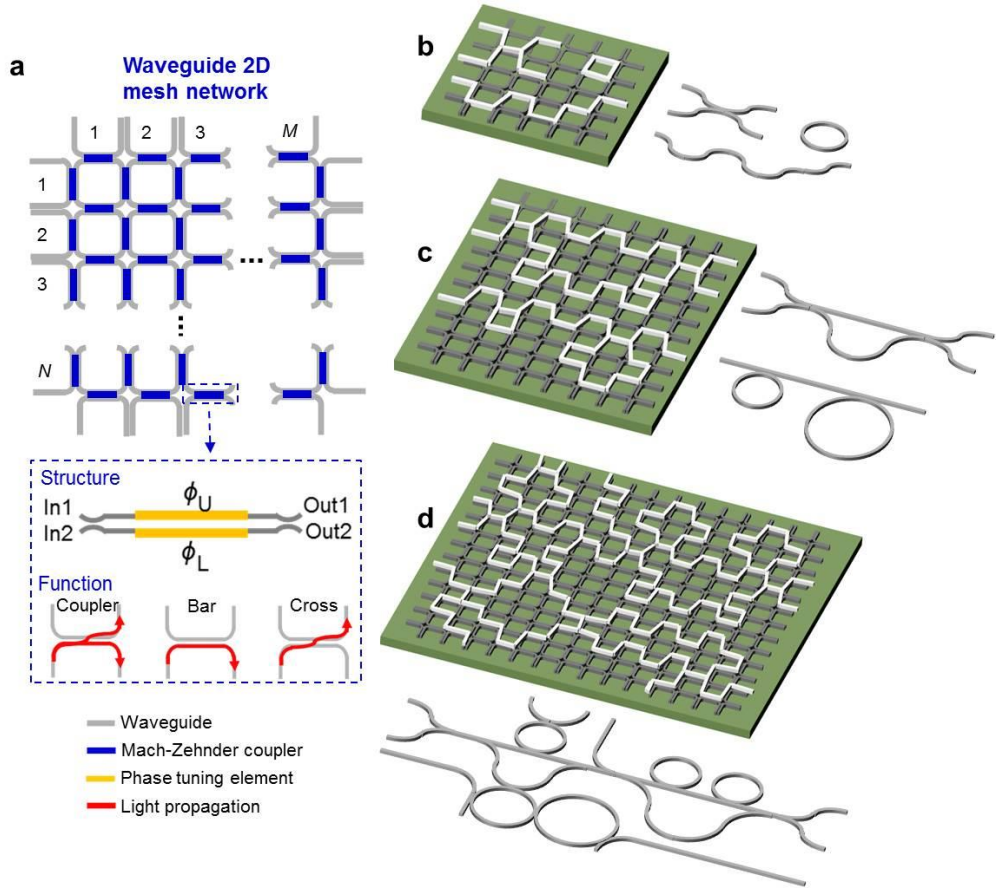

**Figure 1 | Waveguide 2D mesh network. a.** A schematic of a general $M \times N$ waveguide mesh network with Mach-Zehnder couplers being the inter-cell pathways. **b-d.** Examples of implementing various circuit architectures in such mesh networks.

Figure 2a presents a first-demonstrator chip with two mesh cells, fabricated in a commercial $Si_3N_4$ waveguide technology (TriPleX$^{TM}$)[36]. To simplify the fabrication, we use dedicated phase shifters and MZ couplers with a single phase tuning element to perform respectively the effect of $\phi_A$ and $\phi_D$ of MZ couplers with two phase tuning elements. The phase tuning elements are implemented using resistor-based heaters which cause waveguide refractive index change by locally varying the waveguide temperature[10]. On this chip, the phase shifters are found with a full tuning range of 0 to $2\pi$; the power coupling coefficient of the MZ couplers can be tuned very close to the ideal case, i.e., tunable between 0 and 0.99. By programming the values of the phase tuning elements, we demonstrate four distinctively different circuit configurations, including a single-ring notch filter[31], a single-ring Hilbert transformer[32], a dual-ring bandpass filter[33], and a dual-ring delay line[34]. The corresponding settings of the chip and the measurements of the corresponding frequency response shapes that verify the circuit functionalities are depicted in Fig. 2b.

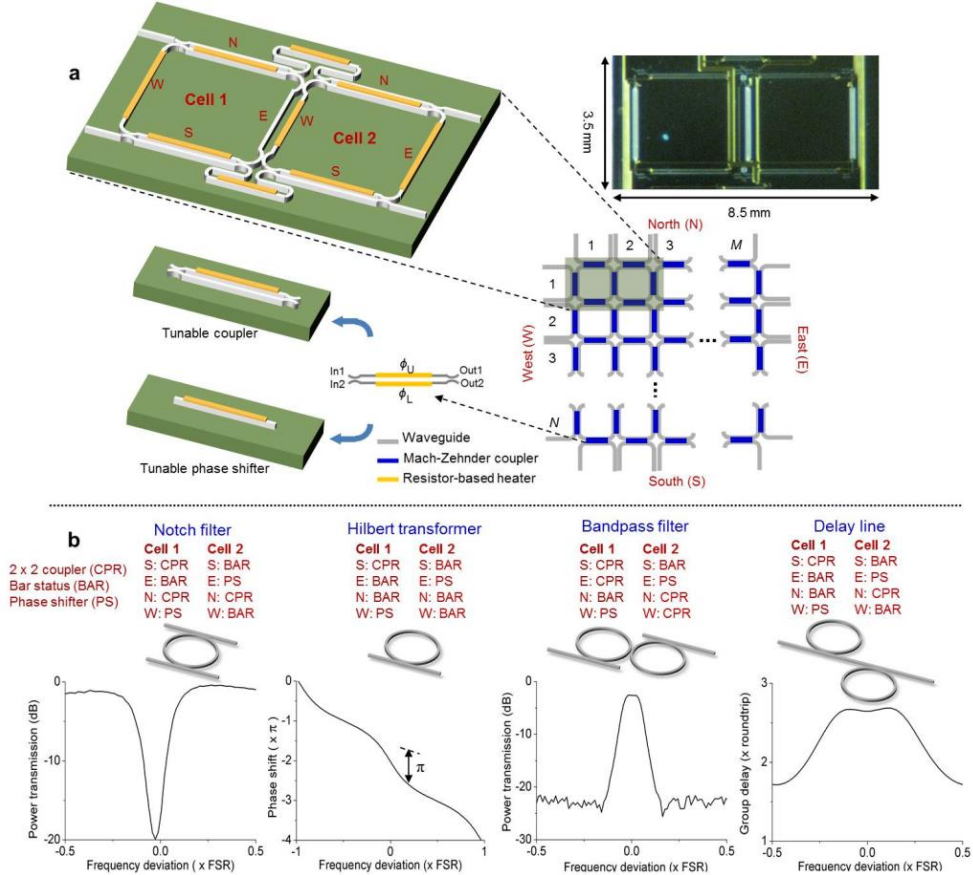

**Figure 2 | A demonstrator chip comprising an elementary 2 × 1 waveguide mesh network. a**, A schematic and a photo of the demonstrator chip fabricated using a commercial $Si_3N_4$ waveguide technology (TriPleX$^{TM}$). **b**, Four different circuit configurations created by programming the phase tuning elements in the chip, and the measurements of their corresponding frequency response shapes.

**RF filter implementation**
Using the demonstrator chip as a programmable photonic signal processor, we implement a new microwave photonic approach of generating RF filters. A schematic of the system and an illustration of the working principle are presented in Fig. 3. Here, an electro-optic modulator is used to create a double-sideband modulation spectrum from an input RF signal under small signal condition[37,38]. The chip is programmed into a circuit comprising a cascade of two ring resonators (as the delay line in Fig. 2b), whose resonance frequency and resonance strength are controllable via phase shifters $\phi_n$ and couplers $\kappa_n$, respectively[35]. We program the two ring resonators such that Ring 1 and Ring 2 have their resonance frequencies in the upper and lower modulation sidebands respectively, and both feature a sharp phase transition and a significant amplitude notch around the resonance frequency and nearly flat phase and amplitude there outside (for simplicity, we consider here only resonance effect for normal dispersions[35]). The equivalent RF responses of the two sidebands after direct detection are depicted alongside, assuming a high-speed photodetector providing sufficient RF bandwidth. The highlighted area exhibits a frequency region where the two RF responses have nearly equal amplitudes and a phase difference of π, in contrast to the equal-phase areas on its two sides. Eventually, these two RF responses add up vectorially at the photodetector output, resulting in a RF filter as illustrated in Fig. 3: a band-stop filter or a band-pass filter, depending on the phase relation between the optical carrier and the sidebands at the modulator output. In practice, a dual-parallel Mach-Zehnder modulator can be used to provide the desired optical spectrum with either in-phase or complementary-phase sidebands (Fig. 3) by controlling the modulator biases[39,40]. Moreover, the programmability of the chip also allows us to implement a RF filter using a conventional microwave photonic approach based on single-sideband modulation[41], where the chip is programmed into an optical filter (e.g. a notch filter as in Fig. 2b). Although also easy to implement, this conventional approach requires an additional processing step to remove one modulation sideband, which increases the system complexity and leads to an extra 3-dB loss in the system gain.

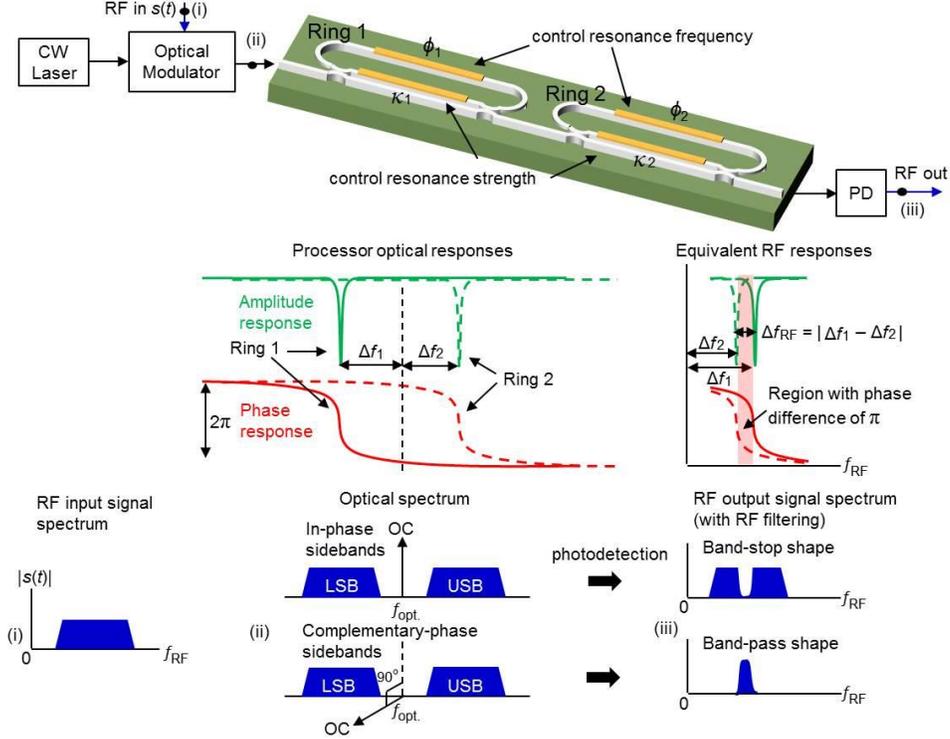

**Figure 3 | RF filter implementation.** A schematic of the microwave photonic system and an illustration of the working principle to implement a RF filter (CW: continuous wave, L/USB: lower/upper sideband, OC: optical carrier, PD: photodetection).

To verify the approach illustrated in Fig. 3, measurements of RF filter responses were performed for both band-stop and band-pass cases. In Figure 4a,b, the measurements show clearly that a band-stop and a band-pass filter can be generated, both having passband-stopband extinctions > 17 dB and passband dispersions < 2 ps/MHz. The fitted curves show that the measured filter shapes are consistent with the theoretical filter transfer function (see Methods). In Figure 4c,d, we demonstrate continuous tuning of the filter center frequency without changes in filter shape. This is performed by controlling the phase shifters ($\phi_1$, $\phi_2$) of the two ring resonators such that $\Delta f_1$ and $\Delta f_2$ (as referred to in Fig. 3) are shifted simultaneously with a constant $\Delta f_{RF} = |\Delta f_1 - \Delta f_2|$. Subject to the frequency periodicity of ring resonators, the maximum frequency coverage of the RF filter equals half of the ring resonator FSR that is 14 GHz in this case. Here, we demonstrate the frequency tuning from 1.6 GHz to 6 GHz (31% of the ring resonator FSR), showing a frequency coverage greater than two octaves. It is worth mentioning that a two-octave frequency coverage in combination with continuous frequency tuning is difficult to achieve with electronic RF filters[42–45]. Besides, our RF filter employs only two tuning elements ($\phi_1$, $\phi_2$) to perform frequency tuning, implying easy control.

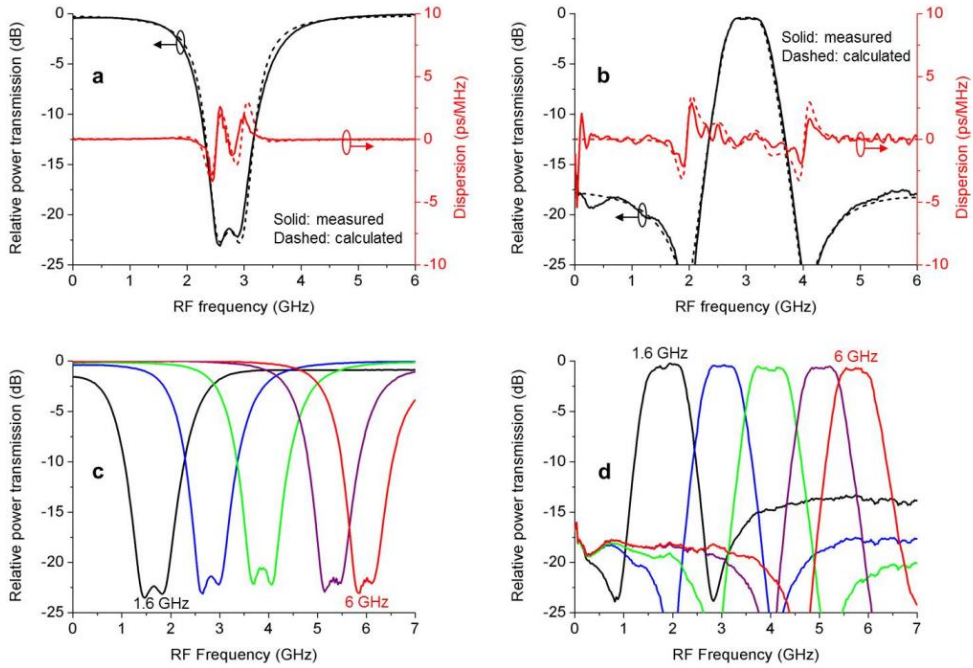

**Figure 4 | RF filter generation and center frequency tuning. a,b,** Measured band-stop and band-pass filter responses and fitted curves of the theoretical filter transfer function. **c, d,** Demonstration of continuous tuning of the filter center frequency.

Next to the continuous frequency tuning, the full control capabilities of the phase shifters ($\phi_1$, $\phi_2$) and couplers ($\kappa_1$, $\kappa_2$) of the two ring resonators also allow for variable passband shaping. Figure 5 presents the measurements of many different filter responses ranging from a 55-dB extinction notch filter to a 1.6-GHz-bandwidth flat-top filter. In Figure 5a,b, we demonstrate variable passband shaping by controlling the phase shifters ($\phi_1$, $\phi_2$). Unlike the operation for the filter center frequency tuning, $\Delta f_1$ and $\Delta f_2$ (as referred to in Figure 3) are shifted independently in this case and the frequency difference between them $\Delta f_{RF} = |\Delta f_1 - \Delta f_2|$ determines the width and ripple of the filter shape. This effect applies to both band-stop and band-pass type of filters. From a practical perspective, however, such as in flat-top filters, it is undesirable to increase the passband ripple when increasing the filter bandwidth. This issue can be addressed by an appropriate setting of the couplers ($\kappa_1$, $\kappa_2$), which is shown in the measurements in Figure 5c. We have achieved wideband flat-top filters with 1-dB-bandwidth of up to 1.6 GHz (11% of the ring resonator FSR or equally 36% of the filter frequency coverage). In addition, when the two couplers ($\kappa_1$, $\kappa_2$) are set with identical coupling coefficients, a passband-stopband extinction of 25 dB can be reached, a measurement of which is shown in Figure 5d. Such RF filters with frequency agility and adjustable bandwidth have great application potential for high-spectrum-efficiency RF technologies such as cognitive radios[46].

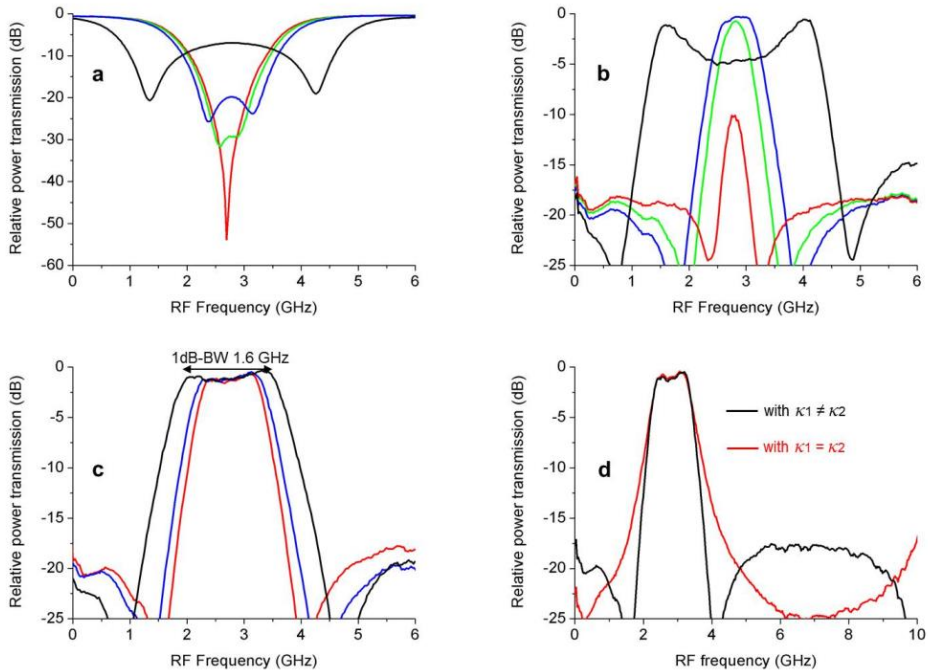

**Figure 5 | Demonstration of variable passband shaping. a,b,** RF filters with different widths and ripples by controlling only $\phi_1$ and $\phi_2$. **c,** Flat-top filters with different bandwidths by controlling $\phi_1$, $\phi_2$, $\kappa_1$, and $\kappa_2$. **d,** A flat-top filter with a passband-stopband extinction of 25 dB by setting $\kappa_1 = \kappa_2$.

**Discussion**

The proper operation of such function-programmable photonic chips relies on the tunability of the MZ couplers, which translates to stringent design requirements for the coupler phase and coupling coefficient tuning range. Our demonstrator chip features good tunability ($\phi = [0, 2\pi]$, $\kappa = [0, 0.99]$), but the complexity of the programmable circuit architectures/functionalities are limited by the small dimension of the mesh network (two mesh cells). However, by scaling up the network dimension, a myriad of functions that are based on more complex circuit architectures are expected to be implementable, such as tapped-delay-line filters, multi-channel (de)multiplexers and cross-connects, high-order coupled resonators, and various combinations of such circuits[35]. With sufficient space on the chip, it is also possible to implement multiple independent functionalities simultaneously, enabling a 'multi-task photonic processor'. On the other hand, however, increasing the network dimension will also raise the difficulties of the calibration and control of the chip, where the required effort for characterizing the possible initial offsets and crosstalks of the tuning elements increases exponentially with the network dimension.

In principle, interferometric waveguide circuits are subject to frequency periodicity[35]. For the proposed waveguide mesh network, the FSRs of its synthesized circuits are inversely proportional to the time delay of the MZ coupler (serving as the basic constituent waveguide in the circuits). Our $Si_3N_4$ waveguide-based demonstrator chip employs MZ couplers with a length of 3450 μm (including a phase shifter/heater section with a length of 2100 μm and two 3-dB directional couplers, each having a length of 675 μm) and a group index of 1.71, based on which the synthesized circuits feature FSRs in the order of tens of GHz. Such FSRs are suitable for RF/microwave signal processing. The capability of larger FSRs will expand the application potential of such photonic signal processor chips, e. g. reconfigurable add-drop multiplexers for optical communications[47]. However, this requires improvements in the waveguide technology to enable further device miniaturization. With this regard, silicon-on-insulator waveguide technology has shown interesting results of tunable MZ couplers with lengths of tens of micrometers[19,20], which promise the realization of the proposed waveguide mesh networks with order-of-magnitude increase of circuit FSRs.

Regarding RF filter implementation, we showed RF filter passbands with a frequency resolution in the order of GHz. This can be further scaled down by means of an according increase of the FSR of the processor chip. However, in the case of a sharper frequency resolution, for instance in the order of tens of MHz as required by mobile communication channels and satellite transponders[1,2], the quality of the CW laser is critical to the filter performance as the filter frequency stability relies largely on the laser linewidth and frequency jitter which may be significant compared to the filter bandwidth. Promisingly, kHz-linewidth lasers have been demonstrated on chip[48], which could be a low-cost solution to address this concern. Other drift-like fluctuations in the system control could be overcome by means of high-resolution tuning and adaptive control algorithms. The chip in this work employs thermo-optical tuning, so temperature control of the chip is required. This tuning mechanism limits the processor programming speed to the range of milliseconds. However, this can be significantly improved by the advancing of the modulator technologies, where the state-of-the-art devices have demonstrated modulation speed in the order of hundreds of picoseconds[49]. Moreover, our RF filter implementation is subject to the principle of a microwave photonic link[37]. This means that the same challenges with respect to system gain, noise figure, and dynamic range also exist. The key to overcome these challenges resides to a large extent in the advancing of optoelectronic components for the conversion between electrical and optical signals. With regards to these properties, promising results have been achieved in the last decade: system gain values larger than 10 dB have been demonstrated, and noise figure values below 6 dB at frequencies beyond 10 GHz, using special, highly-sensitive (low $V_\pi$) modulators and with novel detectors that can handle high currents[50].

## Methods

**Filter transfer function.** The RF filter described in this paper is subject to the principle of a microwave photonic link based on double-sideband modulation and direction detection[37]; however, with the fiber link replaced by a processor chip. The derivations of the overall transfer function of such microwave photonic systems can be found in several previous works[25, 35, 37]. Using the similar derivation steps and assuming small signal condition, the overall transfer function of our RF filter can be expressed as

$$H_{RF}(f) = \sqrt{G(f)} \cdot \sqrt{P_{cl}} \cdot \{C^*_{opt}(-\Delta f_2, \Delta f_1) \cdot [H_{rr\_2}(f - \Delta f_2) + H_{rr\_1}(f + \Delta f_1)]$$
$$+ e^{-j\Delta\theta} \cdot C_{opt}(-\Delta f_2, \Delta f_1) \cdot [H_{rr\_2}(-f - \Delta f_2) + H_{rr\_1}(-f + \Delta f_1)]^*\}$$

where $G(f)$ represents the link gain without the effect of the processor chip; $P_{cl}$ is the optical insertion loss of the processor chip; $\Delta f_n$ is the frequency spacing between the resonance of $n$th ring resonator and the optical carrier; $H_{rr\_n}(f)$ is the complex transfer function of the $n$th ring resonator, $C_{opt}(-\Delta f_2, \Delta f_1) = H_{rr\_2}(-\Delta f_2) + H_{rr\_1}(\Delta f_1)$ is a complex amplitude factor to the optical carrier, and $\Delta\theta$ is the differential phase between the two optical sidebands ($\Delta\theta = 0$ for amplitude modulation and $\pi$ for phase modulation). The transfer function of a ring resonator can be given by

$$H_{rr\_n}(f) = \frac{\sqrt{1-\kappa_n} - \sqrt{L_{rt}} \cdot e^{-j(2\pi f/\Delta f_{FSR} + \phi_n)}}{1 - \sqrt{L_{rt}} \cdot \sqrt{1-\kappa_n} e^{-j(2\pi f/\Delta f_{FSR} + \phi_n)}}$$

where $\kappa_n$ and $\phi_n$ are the coupling coefficient and additional roundtrip phase shift of the $n$th ring resonator, $\Delta f_{FSR}$ the FSR, and $L_{rt}$ the roundtrip loss.

**Processor chip fabrication.** The demonstrator processor chip described in this paper was fabricated using a $Si_3N_4/SiO_2$ waveguide technology (TriPleX$^{TM}$, proprietary to LioniX B.V., The Netherlands) in a CMOS-equipment-compatible process[36]. This waveguide platform allows to modify the waveguide dispersion and polarization properties by changing the cross-section. The waveguides can be optimized to provide extremely low losses, around 0.01 dB/cm (which includes bending losses for a radius of 75 μm) at C-band wavelengths. Tapered facets can be provided to reduce fiber-chip coupling loss to lower than 1 dB/facet. Thermo-optical tuning elements were implemented using Chromium heaters and Gold leads, allowing for a tuning speed in the order of milliseconds.

**Experiments.** The microwave photonic system setup for the RF filter experiment consists of a CW laser with 1 MHz-linewidth (EM4 EM-253-80-557) driven by a high-resolution current source (ILX

Lightwave LDX-3620), a dual-parallel MZ modulator (JDSU 10Gb/s DPMZ), the programmable processor chip, and a photodetector (Discovery semiconductor DSC30S). The RF filter frequency responses presented in Fig. 4 and 5 were obtained by means of the $S_{21}$ network factor measurement using a vector network analyzer (Agilent NA5230A). The configuring of the processor chip was performed by monitoring its optical response, using a high-resolution optical vector analyzer. Programming and calibration of the tuning elements were performed via a dedicated 12-bit-resolution multi-channel voltage supply.

**Acknowledgments**
This research work is enabled by the funding provided from Dutch Agentschap NL IOP project PROMISE2DAY with no. IPD12009 and Australian Research Committee Laureate fellowship with grant no. FL13010041.

**Author contributions**
L.M.Z. conceived and performed the experiment, and designed the photonic chip; C.G.H.R. and A.J.L. led the research work and organized the experiment setup; M.H. co-designed and fabricated the photonic chip; K.-J.B., C.G.H.R., M.H. and A.J.L. co-wrote the paper.

**Additional information**
The authors declare no competing financial interests.